# Characterizing Habitable Exo-Moons

L. Kaltenegger
*Harvard University, 60 Garden Street, 02138 MA, Cambridge USA*
*Email: lkaltene@cfa.harvard.edu*

**Abstract**
We discuss the possibility of screening the atmosphere of exomoons for habitability. We concentrate on Earth-like satellites of extrasolar giant planets (EGP) which orbit in the Habitable Zone (HZ) of their host stars. The detectability of exomoons for EGP in the HZ has recently been shown to be feasible with the Kepler Mission or equivalent photometry using transit duration observations. Transmission spectroscopy of exomoons is a unique potential tool to screen them for habitability in the near future, especially around low mass stars. Using the Earth itself as a proxy we show the potential and limits of spectroscopy to detect biomarkers on an Earth-like exomoon and discuss effects of tidal locking for such potential habitats.

**INTRODUCTION**

Transiting planets are present-day "Rosetta Stones" for understanding extrasolar planets because they offer the possibility of characterizing giant planet atmospheres (see. e.g. Tinetti et al. 2007, Swain et al. 2008) and should provide an access to biomarkers in the atmospheres of Earth-like bodies (Kaltenegger & Traub 2009, Deming et al. 2009, Ehrenreich et al. 2006), once they are detected. Extrasolar giant planets might have "exomoon" satellites that could be detected using transit time duration measurements (Sartoretti & Schneider 1999, Agol et al. 2005, Holman & Murray 2005, Kipping 2009), lightcurve distortions (Szabo et al. 2006), planet-moon eclipses (Cabrera & Schneider 2007), microlensing (Han 2008), pulsar timing (Lewis et al. 2008) and distortions of the Rossiter-McLaughlin effect of a transiting planet (Simon et al. 2009). A detailed study (Kipping et al. 2009) using transit time duration measurements found that exomoons around Extrasolar Giant Planets (EGP) in the Habitable Zone (HZ) of their host star down to 0.2 Earth masses ($M_E$) may be detectable with the Kepler Mission (Borucki et al. 1997) or equivalent photometry, which translates into 25000 stars within Kepler's field-of view that can be screened for Earth-mass moons. This leads to the question of whether we could screen such Earth-mass exomoons remotely for habitability.

If the viewing geometry is optimal, i.e. the exomoon is close to the projected maximum separation, the transit of an exomoon is comparable to the transit of a planet the same size around the star (Kaltenegger & Traub 2009), if the projected semimajor axis is larger than the stellar radius. In such a case transmission spectroscopy is a unique potential tool to screen exomoons for habitability in the near future. Section 2 discusses the stability and geometry of potentially habitable exomoons, section 3 detectable spectral features of Earth-analog exomoons, section 4 describes our model, section 5 presents and section 6 discusses the results.

**2. A UNIQUE GEOMETRY**

Several groups have investigated the long-term dynamical stability of hypothetical satellites and moons up to Earth masses orbiting EGP (Cassidy et al. 2009, Barnes & O'Brian 2002, Domingos et al. 2006, Scharf 2006, Holman & Wiegert 1999) and have shown that Earth-mass moons could exist and be stable for more than 5Gyr around EGP. The maximum stable distance of a satellite from its planet is given by the Hill Radius $R_H$ that depends on the distance of the planet to its host star $a_p$, as well as the mass of the planet $M_p$ and the star $M_{star}$ respectively. In detail, the critical semimajor axis beyond which the satellite would not be stable also depends on the planet's eccentricity $e_p$ as well as the satellite's eccentricity $e_{Sat}$ and whether it is in a retrograde ($a_{Sat} = a_{eR}$) or prograde ($a_{Sat} = a_{eP}$) orbit (see Domingos et al. 2006 for details), with additional stable regions further out for periodic orbits that are not considered here (see e.g. Henon 1969). The critical distance depends only weakly on the mass ratio (Holman & Wiegert 1999), so we can safely use the approximation of Domingos et al. (2006) which was derived for a mass ratio of $10^{-3}$.

$$R_H = a_p (M_p / 3M_{Star})^{1/3} \qquad (1)$$

$$a_{eR} \approx 0.9309 R_H (1 - 1.0764 e_p - 0.9812 e_{Sat}) \qquad (2)$$

$$a_{eP} \approx 0.4895 R_H (1 - 1.0305 e_p - 0.2738 e_{Sat}) \qquad (3)$$

We calculate the critical semimajor axis for an Earth-like exomoon, for retrograde and prograde orbits around a Jupiter mass ($M_J$) and a 13 $M_J$ EGP in the Habitable Zone of their host star. We concentrate on small stars because the transit probability $p$ increases with decreasing stellar mass, which favors M dwarfs. $p \approx R_s/a_p$ where $R_s$ the radius of the star and $a_p$ the semi-major axis of the planet. The distance $a_{HZ}$ is defined here as the 1AU-equivalent, where a planet would receive the same stellar flux as the Earth, $a_{HZ} = 1\ AU\ (L_{star}/L_{Sun}\ S_{eff})^{0.5}$, where $L$ is the luminosity of the star and the normalized solar flux factor $S_{eff}$ takes the wavelength dependent intensity distribution of the spectrum of different spectral classes into account, it is set to 1 for the Sun and 1.05 for cool stars (Kasting et al. 1993).

## 3. REMOTELY SCREENING ATMOSPHERES FOR HABITATS

Currently several exoplanets are known to have a minimum mass below 10 $M_E$ (Mayor et al. 2009). This mass limit is usually considered to separate terrestrial from giant planets, the latter having a significant fraction of their mass in a $H_2$-He envelope. Planets and satellites without a massive $H_2$-He envelope are potentially habitable because they may have suitable temperatures and pressures at any solid surface to support liquid water. On Earth, some atmospheric species exhibiting noticeable spectral features in the planet's spectrum result directly or indirectly from biological activity: the main ones are $O_2$, $O_3$, $CH_4$, and $N_2O$. $CO_2$ and $H_2O$ are in addition important as greenhouse gases in a planet's atmosphere and potential sources for high $O_2$ concentration from photosynthesis (see Kaltenegger et al. 2009a, 2009b for details).

A habitable moon would need to be able to retain its volatiles. This depends on its mass, the charged particle flux it receives, whether it maintains a magnetosphere and on its position with respect to the planet's magnetosphere, among other factors. This leads to a lower mass limit between 0.12 to 0.23 $M_E$ (Williams et al. 1997, Kaltenegger et al. 2000) above which moons can potentially be Earth-analog environments if their planet orbits within the HZ.

Our search for signs of life is based on the assumption that extraterrestrial life shares fundamental characteristics with life on Earth, in that it requires liquid water as a solvent and has a carbon-based chemistry (see e.g. Brack 1993), uses the same input and output gases and exist out of thermodynamic equilibrium (Lovelock 1975). 'Biomarkers' is used here to mean detectable chemical species, whose presence at significant abundance strongly suggests a biological origin (e.g. $CH_4 + O_2$, or $CH_4 + O_3$, $N_2O$). 'Bioindicators' (e.g. $H_2O$, $CH_4$) are indicative of biological processes but can also be produced abiotically. It is their abundances, and their detection in the context of both other atmospheric species, and the properties of the star and the planet that points toward a biological origin (Kaltenegger & Selsis 09).

## 4. MODEL DESCRIPTION

The area of the stellar disk that is blocked by a transiting planet is given by $\pi R^2(\lambda)$, where $R(\lambda) = R_p + h(\lambda)$, $R_p$ is the radius of the planet at the base of the atmosphere, and $h(\lambda)$ is the effectively opaque height of the atmosphere at that wavelength. $h(\lambda)$ can be as large as about 50 km for strong spectral features in the Earth's atmosphere (Kaltenegger & Traub 2009). During a transit, the relevant signal of atmosphere absorption is $N(sig) = N(cont) - N(line)$, where $N(cont)$ is the number of potentially detectable photons in the interpolated continuum at the location of a spectral feature and over the wavelength band of that feature, and $N(line)$ is the number of detected photons in the band. The noise $N(noise)$ in an idealized case is the fluctuation in the total number of detected photons from the star $N(tot)$ in the same wavelength band, so $N(noise) = N^{1/2}(tot)$, ignoring all other noise contributions. The total number of photons detected from the star during a transit $N(sig)$, in a given spectral range is $N(sig) = N(tot)\, 2R_p h(\lambda)/R_s^2$. Thus the SNR for detecting a given spectral feature is $SNR = N^{1/2}(tot)\, 2R_p h(\lambda)/R_s^2$.

Model Earth spectra are calculated with the Smithsonian Astrophysical Observatory code developed to analyze balloon-borne spectra of the stratosphere (see e.g. Traub & Stier, 1976). The spectral line database includes the large HITRAN compilation and other sources (Rothman et al. 2009; Yung & DeMore 1999). The far wings of pressure-broadened lines can be non-Lorentzian at around 1,000 times the line width and beyond; therefore, in some cases ($H_2O$, $CO_2$, $N_2$) we replace line-by-line calculation with measured continuum data in these regions. Aerosol and Rayleigh scattering are approximated by applying empirical wavelength power laws (Cox 2000). Atmospheres are constructed from standard models that are discretized to appropriate atmospheric layers. Clouds are represented by inserting continuum-absorbing/emitting layers at appropriate altitudes. For the transit spectra, we assume that the light paths through the atmosphere can be approximated by incident parallel rays, bent by refraction as they pass through the atmosphere, and either transmitted or lost from the beam by single scattering or absorption. We model the Earth's spectrum using its spectroscopically most significant molecules, $H_2O$, $O_3$, $O_2$, $CH_4$, $CO_2$, CFC-11, CFC-12, $NO_2$, $HNO_3$, $N_2$ and $N_2O$, where $N_2$ is included for its role as a Rayleigh scattering species. Our line-by-line radiative transfer code for the Earth has been validated by comparison to observed reflection, emission and transmission spectra (Woolf et al. 2002; Christensen & Pearl 1997, Kaltenegger et al. 2007, Irion et al. 2002, Kaltenegger & Traub 2009). For this paper we use an Earth model atmosphere, namely the US Standard Atmosphere 1976 (COESA 1976) spring-fall model to derive the strength of the detectable feature.

## 5. RESULTS

Table 1 shows the stellar parameters, the 1AU-equivalent distance $a_{HZ}$, the planet's orbital period for a 1 Jupiter mass $P_{1J}$ with an Earth-mass exomoon and the corresponding transit duration $T_{1J}$. The period and transit duration is reduced by less than 10 % for a 13 $M_J$ EGP orbiting the smallest host star (M9). $n_T$ denotes the number of transits per Earth year, and $\sigma$ the total number of hours the planet can be observed in transits per Earth year.

Table 2 shows the maximum orbital separation $a_e$ of an exomoon from its planet in stellar radii. We assume circular orbits and calculate the corresponding orbital periods $P_e$ to assess if the exomoon transits can occur uncontaminated by the planet's transit. A large projected angular separation of the moon from its planet leads to a high probability of an uncontaminated exomoon transit, which could be detected and observed individually. In all cases $a_e$ is equivalent to several stellar radii, therefore the transit of an exomoon can potentially be observed individually for the best viewing geometry, where the exomoon either trails or precedes its planet. Column 2 show the contrast ratio $f_{p100km}$ of an opaque 100 km high absorption feature in the atmosphere of the transiting body (see Fig.1). The absolute contrast ratio of each individual feature is given by multiplying $f_{p100km}$ with the effective height $h(\lambda)/100$.

Fig. 1 shows the absorption depth and main detectable features in low resolution in the 0.3 – 4.0 μm range: $O_3$, $H_2O$, $CO_2$, $CH_4$, and potentially $O_2$, in order of decreasing strength as well as in the 4 - 20 μm range: $CO_2$, $O_3$, $CH_4$, $H_2O$, and $HNO_3$, in order of decreasing strength.

**5.1. SNR for Exomoons**

We calculate the achievable SNR for primary eclipse measurements for the closest stars. The value of *N(tot)* is calculated assuming an effective temperature of 5770 K for the Sun and the values given in Table 1 for M stars. The photon rate from each star is computed assuming that it is a black body, which is a crude approximation for late type stars and leads to an overestimation of the SNR for some of the shorter wavelength transiting spectral features.

The number of detected photons is computed assuming a 6.5-m diameter telescope like the James Webb Space Telescope (JWST), a net efficiency of 0.15 electrons/photon and only photon noise. Here we assume an observing mode which allows observing nearby stars with JWST.

Table 3 lists the strongest features from the effective height spectrum in Fig. 1, including central wavelength, full width at half maximum, and the average effective height of the feature. Column 5 gives the number of transits needed to achieve a SNR of 3 for the closest stars of each subclass – the most interesting targets - for each feature for an Earth-like satellite in the HZ of the Sun, and columns 6 – 10 give the number of transits for these features for an Earth-like satellite in the HZ of M0 – M9 dwarf stars for the closest stars (Reid et al. 2005, Reid in prep). Gl 559 A ($\alpha$ centauri A, G2V) at 1.34pc, Gl 887 at 3.29pc (M0.5), Gl 411 at 2.54pc (M2), Gl 551 at 1.30pc (M5.5), Gl 473 Bat 4.39pc (M7), and Denis1048 at 4.03pc (M9), making $\alpha$ Centauri A, Gl 551 and Denis1048 the best candidates to search for habitable exomoons in order of integration time needed to achieve a SNR of 3. Table 4 shows the integration time needed in transit to observe features with a SNR of 3 for stars at a distance of 10pc.

**6. DISCUSSION**

Retrograde motions of moons are rare in the Solar System, most caused by capturing. In retrograde motion tidal forces lead to a slow decay of the orbit until the moon reaches the Roche limit. Extrapolating from our own Solar System we would expect moons in retrograde orbits to be rare.

The apparent radius of the Earth-analog varies by a maximum of 50 km due to absorption, less than 1%. Table 3 shows that multiple transits are needed for the physical parameters of the closest M stars as well as the closest Sun-like star to detect atmospheric features on an Earth-analog exomoon in transit, except for $\alpha$ Centauri A. At a fixed distance from the observer it is easier to detect exomoons around the lowest mass stars because of the favorable contrast of the star to the planet (see table 4). Note that measurements indicate that $\alpha$ Centauri A and proxima Centauri are unlikely to harbor an EGP (Endl & Kürster 2008, D. Fischer, private communication 2009). The frequency of giant planet within the HZ is unknown and might be low for M stars (see, e.g, Endl et al. 2006).

In the best case scenario atmospheric $H_2O$, $CO_2$ and $O_3$ features in the IR could be detected in one Earth year for transiting habitable exomoons around M5 to M9 stars for a distance up to 10pc, if such exomoons exist. Note that absorption lines in the stellar spectrum (see e.g. Mohanty, S. et al. 2005 for M star spectra) can significantly lower the SNR at certain wavelengths. In addition to the time shown in table 3 and 4, a minimum of the same time of out of transit observation of the star is needed to make these measurements. Under idealized observing conditions, one transit observation of an Earth-like body around $\alpha$ Centauri A (Cameron et al. 2009) with JWST would potentially let us screen it for habitability, making it an excellent target for the search for habitable exomoons.

**6.1 M Star Atmospheres**

We do not adjust the spectrum of the exomoons photochemistry in these calculations to the detailed incident spectral distribution of M stars. Several groups have shown that an Earth-analog spectrum around small stars is in first order comparable to the Earth's absorption strength (see e.g. Segura et al. 2005) but for a potential increase in the abundance of $CH_4$. Increased methane abundance would make the detection of the 7.7 µm $CH_4$ feature feasible, which is the limiting case to characterize the major atmospheric features in an Earth-analog feature in transit.

To assess the effect of the satellites on the depth of the absorption lines, we calculate the scale height for such objects from 0.2 to 10 $M_E$. $N(sig)$ is proportional to $h(\lambda)$ and $R_p$. In first order $h(\lambda)$ is proportional to the atmospheric scale height $H$, which depends on the mean molecular weight of the atmosphere $\mu_{mol}$, the body's gravity $g$, the temperature $T$ and the Boltzmann constant $k$. $H = k T / (\mu_{mol} g)$. $R_p$ is proportional to the $M_p^{0.3}$ for small objects and $M_p^{0.27}$ for Super-Earths (Valencia et al. 2006) while the gravity is proportional to the mass of the planet divided by the radius squared, therefore $N(sig)$ is proportional to $M_p^{-0.1}$ for small, and $M_p^{-0.19}$ for massive rocky satellites respectively. For a mass range from 0.2 to 10 $M_E$, that translates into a very small correlation of line depth with the mass of the body and a maximum of 35% increase for a 0.2 $M_E$ body and 20% decrease for a 10 $M_E$ Super-Earth, which we do not consider here.

**6.2 Tidally Locked Planets and Moons**

Close by M stars are ideal candidates for detection and subsequent characterization of potentially habitable exomoons because of the small distance of their HZ, which increases the transit probability as well as the transit frequency per observation time. In addition habitable exomoons around M stars would be tidally locked to their planet, not to their host star, removing the problem of a potential freeze out of the atmosphere on the dark side of an Earth-like exomoon, which has been discussed as a concern for tidally locked planets around M stars (Scalo et al. 2007, Joshi et al. 1997).

**6.3. Comparison to combined Emergent versus Transmission Spectroscopy**

The difference in surface area and atmospheric scale height of an EGP versus an Earth-size exomoon limits the detectability of spectral signatures from a potentially habitable exomoon if the two light sources can not be spatially separated. Here we use Jupiter as an example for an EGP of small mass and radius to compare the strength of the signals, assuming a similar Bond albedo. This calculation quantitatively shows the order of magnitude of the signal strength difference as a first assessment of the detectability of spectral features from the satellites atmosphere embedded in the EGP spectra. Initial models that include model spectra for EGP in the HZ have been published and depend strongly on the EGP atmosphere model (Williams & Knacke 2004).

The reflected as well as emitted signal for emergent spectra depends on the surface area of the planet. Jupiter has about 11 times Earth's radius, 2.4 times its gravity and the mean molecular weight of its atmosphere is about 2.22 g/mol versus Earth's 28.97 g/mol. These values lead to an emergent EGP spectra signal more than 120 times stronger than Earth's due to the increased surface area ($R_J \approx 11\ R_E$). The change in eclipse depth across spectral lines in transmission is in first order proportional to the scale height - about 7 times Earth's for Jupiter - times the planetary radius leading to an about 80 times stronger signal. Even though the features of an EGP and an Earth-like exo-moon are expected to be different, the two order magnitude difference in feature strength would make it extremely difficult to detect a moon's atmospheric spectral feature embedded in an EGP planet's spectrum if the two bodies can not be spatially resolved.

**7. CONCLUSIONS**

Habitable-zone exomoons may be detected in the near future with missions like Kepler and could be orbiting their planet at a distance that allows for spatially separate transit events. In that case transmission spectroscopy of Earth-like exomoons is a unique potential tool to screen them for habitability in the near future, especially for M stars. Spatially separating the exomoons from their parent planet improves their detectability because their absorption signature are about two orders of magnitude lower than the absorption features of en EGP spectra. We show that the number of transits needed under idealized conditions and viewing geometry is feasible using JWST for the sample of the closest M stars as well as the closest G star, α Centauri A – under the assumption that these stars have Earth-like exomoons orbiting an EGP in the Habitable Zone of their host star. Here we assume an observing mode which allows observing nearby stars with JWST.

The closest M stars are ideal candidates for such observations because of the small distance of their Habitable Zone, which increases the transit probability as well as the transit frequency per observation time. The contrast ratio of atmospheric absorption features of an Earth-analog transiting an M9 star is two orders of magnitude lower than the same object transiting a Sun-like star (see Fig.1 and table 2).

In addition habitable exomoons around M stars would be tidally locked to their planet and not to their host star, which removes the problem of a consequent freeze out of the atmosphere.

ACKNOWLEDGEMENT

Special Thanks to D. Kipping, J. McDowell and G. Torres for stimulating discussions. LK gratefully acknowledges the Harvard Origins of Life Initiative and the NASA Astrobiology Institute.

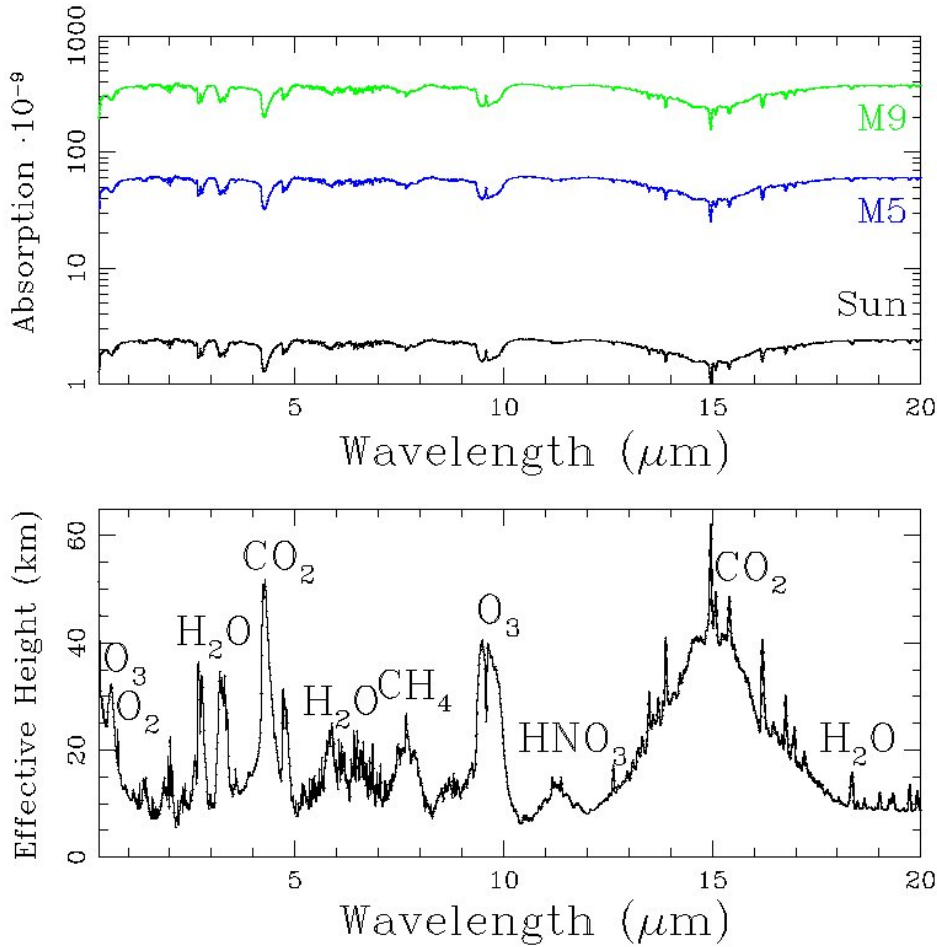

**Fig.1: (top panel) Contrast ratio of the absorption features of an Earth-like atmosphere calculated for the transit of an exomoon. The absolute value depends on the host star (see Table 1 for values of $f_{p100km}$). (lower panel) Effective height of the absorbing atmosphere.**

**Table 1: Stellar parameters, and derived Extrasolar Giant Planet parameters**

| Star | T | R | Mass | a(HZ) | $P_{1J}$ | $T_{1J}$ | $n_T$ | σ |
|---|---|---|---|---|---|---|---|---|
| | K | $R_{sun}$ | $M_{sun}$ | AU | Hr | hr | $yr^{-1}$ | hr/yr |
| Sun | 5770 | 1.00 | 1.00 | 1.000 | 8760 | 12.96 | 1.0 | 12.9 |
| M0 | 3800 | 0.62 | 0.60 | 0.262 | 1514.2 | 5.31 | 5.8 | 30.7 |
| M1 | 3600 | 0.49 | 0.49 | 0.183 | 975.3 | 3.87 | 9.0 | 34.8 |
| M2 | 3400 | 0.44 | 0.44 | 0.148 | 751.1 | 3.30 | 11.7 | 38.6 |
| M3 | 3250 | 0.39 | 0.36 | 0.120 | 602.5 | 2.91 | 14.5 | 42.3 |
| M4 | 3100 | 0.26 | 0.20 | 0.072 | 380.5 | 2.02 | 23.0 | 46.6 |
| M5 | 2800 | 0.20 | 0.14 | 0.046 | 228.5 | 1.48 | 38.4 | 56.7 |
| M6 | 2600 | 0.15 | 0.10 | 0.029 | 138.1 | 1.05 | 63.5 | 66.5 |
| M7 | 2500 | 0.12 | 0.09 | 0.022 | 93.6 | 0.76 | 93.6 | 71.3 |
| M8 | 2400 | 0.11 | 0.08 | 0.017 | 67.7 | 0.65 | 129.6 | 84.4 |
| M9 | 2300 | 0.08 | 0.08 | 0.012 | 41.5 | 0.41 | 211.1 | 86.8 |

**Table 2: Maximum orbital separation in stellar radii and period of an exomoon in hrs[1]**

| Star | $f_{p100km}$ | $a_{eP\_1J}$ | $a_{eP\_13J}$ | $P_{eP}$ | $a_{eR\_1J}$ | $a_{eR\_13J}$ | $P_{eR}$ |
|---|---|---|---|---|---|---|---|
| | | $R_{Star}$ | $R_{Star}$ | Hr | $R_{Star}$ | $R_{Star}$ | hr |
| Sun | 2.63E-09 | 7.2 | 16.9 | 1732.1 | 13.7 | 32.1 | 4542.5 |
| M0 | 6.85E-09 | 3.6 | 8.5 | 299.6 | 6.8 | 16.1 | 785.8 |
| M1 | 1.10E-08 | 3.4 | 8.0 | 193.0 | 6.5 | 15.2 | 506.2 |
| M2 | 1.36E-08 | 3.2 | 7.5 | 148.7 | 6.0 | 14.2 | 389.9 |
| M3 | 1.73E-08 | 3.1 | 7.3 | 119.3 | 5.9 | 13.8 | 312.8 |
| M4 | 3.90E-08 | 3.4 | 8.0 | 75.4 | 6.5 | 15.3 | 197.8 |
| M5 | 6.59E-08 | 3.2 | 7.4 | 45.3 | 6.0 | 14.2 | 118.9 |
| M6 | 1.17E-07 | 3.0 | 7.1 | 27.4 | 5.7 | 13.5 | 72.0 |
| M7 | 1.83E-07 | 2.9 | 6.9 | 18.6 | 5.5 | 13.0 | 48.8 |
| M8 | 2.18E-07 | 2.6 | 6.0 | 13.5 | 4.9 | 11.5 | 35.3 |
| M9 | 4.12E-07 | 2.5 | 6.0 | 8.3 | 4.8 | 11.4 | 21.7 |

**Table 3:** The integer number of uncontaminated transits ($N_T$) needed to detect the major spectroscopic features with a SNR of 3 for a transiting Earth-like exomoon, a 6.5-m space based telescope, an efficiency of 0.15 and the closest stars per stellar subtype.

| 6.5-m telescope | | | | $N_T$ for closest stars (SNR 3) | | | | | |
|---|---|---|---|---|---|---|---|---|---|
| Feature | λ | Δλ | h(λ) | G2V | M0V | M2V | M5V | M7V | M9V |
| | μm | μm | km | 1.34pc | 3.29pc | 2.54pc | 1.30pc | 4.39pc | 4.03pc |
| $O_3$ | 0.6 | 0.15 | 10 | 1 | 3 | 3 | 2 | 33 | 50 |
| $H_2O$ | 1.9 | 0.2 | 5 | 1 | 10 | 6 | 1 | 14 | 13 |
| $CO_2$ | 2.8 | 0.1 | 20 | 1 | 3 | 2 | 1 | 3 | 2 |
| $H_2O$ | 3.3 | 0.25 | 20 | 1 | 1 | 1 | 1 | 2 | 1 |
| $CH_4$ [2] | 7.7 | 0.7 | 7 | 4 | 38 | 21 | 3 | 31 | 24 |
| $O_3$ | 9.8 | 0.7 | 30 | 1 | 4 | 2 | 1 | 3 | 2 |
| $CO_2$ | 15.2 | 3.0 | 25 | 1 | 4 | 2 | 1 | 3 | 3 |

---

[1] The orbital period of an exomoon at maximum separation does not depend on the mass of its host planet see equation (1)

[2] See discussion on the abundance of $CH_4$ for M-stars that makes the concentration shown a lower limit

**Table 4:** Exposure time in hrs of transits needed to detect the major spectroscopic features with a SNR of 3 for a transiting Earth-like exomoon, using a 6.5-m space based telescope with an efficiency of 0.15 and stars at 10pc.

| 6.5-m telescope | | | | exposure time (hr) in transit, stars at 10pc (SNR 3) | | | | | |
|---|---|---|---|---|---|---|---|---|---|
| Feature | λ | Δλ | h(λ) | G2V | M0V | M2V | M5V | M7V | M9V |
| | μm | μm | km | 10 pc | 10 pc | 10 pc | 10 pc | 10 pc | 10 pc |
| $O_3$ | 0.6 | 0.15 | 10 | 41.6 | 142.8 | 149.5 | 130.8 | 123.1 | 119.0 |
| $H_2O$ | 1.9 | 0.2 | 5 | 522.7 | 469.8 | 308.6 | 105.1 | 52.4 | 30.4 |
| $CO_2$ | 2.8 | 0.1 | 20 | 165.1 | 126.2 | 78.1 | 23.4 | 10.7 | 5.8 |
| $H_2O$ | 3.3 | 0.25 | 20 | 99.2 | 72.5 | 44.1 | 12.8 | 5.7 | 3.0 |
| $CH_4$ | 7.7 | 0.7 | 7 | 2889.6 | 1839.6 | 1065.3 | 278.0 | 114.9 | 57.0 |
| $O_3$ | 9.8 | 0.7 | 30 | 313.8 | 195.7 | 112.5 | 28.9 | 11.8 | 5.8 |
| $CO_2$ | 15.2 | 3.0 | 25 | 346.8 | 211.2 | 120.4 | 30.4 | 12.3 | 6.0 |